Comment on
"Investigation of simplified thermal expansion models for compressible Newtonian fluids applied to nonisothernal plane Couette and Poiseuille flows" by S. Bechtel, M. Cai, F. Rooney and Q. Wang [Physics of Fluids, 2004, Vol. 16, pp. 3955-3974]

Asterios Pantokratoras
Associate Professor of Fluid Mechanics
School of Engineering, Democritus University of Thrace,
67100 Xanthi – Greece
e-mail:apantokr@civil.duth.gr

In the above paper by Bechtel, Cai, Rooney and Wang, Physics of Fluids, 2004, 16, 3955-3974 six different theories of a Newtonian viscous fluid are investigated and compared, namely, the theory of a compressible Newtonian fluid, and five constitutive limits of this theory: the incompressible theory, the limit where density changes only due to changes in temperature, the limit where density changes only with changes in entropy, the limit where pressure is a function only of temperature, and the limit of pressure a function only of entropy. The six theories are compared through their ability to model two test problems: (i) steady flow between moving parallel isothermal planes separated by a fixed distance with no pressure gradient in the flow direction (Couette flow), and (ii) steady flow between stationary isothermal parallel planes with a pressure gradient (Poiseuille flow). The authors found, among other, that the incompressible theory admits solutions to these problems of the plane Couette/Poiseuille flow form: a single nonzero velocity component in a direction parallel to the bounding planes, and velocity and temperature varying only in the direction perpendicular to the planes.

For the combination of incompressible theory and Poiseuille flow the authors presented dimensionless temperature and dimensionless velocity profiles for Brinkman number 2 (Br=2) and E=0, 3, 10 and 37 (page 3967 in their work). The Brinkman number and the viscosity parameter E are defined as (table II in their work)

$$Br = \frac{\beta h^2}{(k\vartheta_w \mu_w)^{1/2}} \qquad (1)$$

$$E = \frac{\varepsilon}{\vartheta_w} \qquad (2)$$

where β is the pressure gradient, h is the distance between the planes, k is the fluid thermal conductivity, $\theta_w$ is the absolute temperature of the planes, $\mu_w$ is the fluid dynamic viscosity at the planes and ε is a constant included in the following equation which gives the fluid viscosity as a function of temperature (equation 49 in their work)

$$\mu(\vartheta) = \mu_w \exp\left[\varepsilon\left(\frac{1}{\vartheta} - \frac{1}{\vartheta_w}\right)\right] \qquad (3)$$

However, the temperature and velocity profile for E=37 included in their figure 12 does not exist in reality. Our argument is based on the following facts. In two recent papers Costa and Mecedonio (2003, 2005) mention that in a steady–state fully developed Poiseuille flow of an incompressible fluid with temperature-dependent viscosity there is a dimensionless parameter G which is an important criterion for this flow. If G>G$_{crit}$ the system does not admit a solution, whereas when G<G$_{crit}$ the system has two solutions, one of which (the solution with greater temperature) may be unstable. The existence of a critical quantity which characterizes the flow in ducts for fluids with temperature-dependent viscosity including viscous heating was suggested by Grundfest (1963). A very good review on this matter is given by Sukanek and Laurence (1974). The parameter G is defined as

$$G = \frac{c\beta^2 h^4}{k\mu_w} \qquad (4)$$

where h is the half distance between the planes and c is the viscosity parameter used by Costa and Macedonio included in the following equation which gives the fluid viscosity as a function of temperature

$$\mu(\vartheta) = \mu_w \exp[-c(\vartheta - \theta_w)] \tag{5}$$

The critical value of G is 5.64 ($G_{crit}$=5.64, Costa and Macedonio, 2003, page 550). In equations (4) and (5), which concern the work of Costa and Macedonio, we used the symbols used by Bechtel et al. (2004) to avoid confusion and to make some comparisons easier. The viscosity parameter E and the Brinkman number Br of Bechtel et al. (2004) are equivalent (not exactly the same) to quantities c and G used by Costa and Macedonio. The only difference between the Poiseuille problem treated by Bechtel et al. (2004) and that treated by Costa and Macedonio (2003) is the different function for fluid viscosity. Taking into account this fact it is natural to assume that a similar criterion exists for the problem treated by Bechtel et. al. In order to find this criterion we solved the equations (73a) and (73b) given by Bechtel et al. with a finite difference method.

The equations (73a) and (73b) represent a two-dimensional parabolic flow. Such a flow has a predominant velocity in the streamwise coordinate (unidirectional flow) which in our case is the direction along the planes. The equations were solved using the finite difference method of Patankar (1980). The solution procedure starts with a known distribution of velocity and temperature at the channel entrance (x=0) and marches along the plates. These profiles were used only to start the computations and their shape had no influence on the results which were taken far downstream. At the channel entrance the temperature and velocity were taken uniform with a very small value. At each downstream position the discretized equations are solved using the tridiagonal matrix algorithm (TDMA). As x increases the successive velocity profiles become more and more similar and the same happens with temperature profiles. The solution procedure stops at the point where the successive velocity and successive temperature profiles become identical (fully developed flow both hydrodynamically and thermally). The forward step size Δx was 0.01 mm and the lateral grid cells 500. The results are grid independent. The parabolic solution procedure is a well known solution method and has been used extensively in the literature. It appeared for the first time in 1970 (Patankar and Spalding, 1970) and has been included in classical fluid mechanics textbooks (see page 275 in White 1991). In the solution procedure μ has been considered as function of temperature (equation 3, present work). The solution procedure has been applied successfully in a similar problem (Pantokratoras, 2006). A detailed description of the solution procedure, with variable thermophysical properties, may be found in Pantokratoras (2002).

Before applying our solution procedure to the problem of Bechtel at al. (2004) we applied it to the problem of Costa and Macedonio (2003) and we found that the critical value of G is 5.64. For values of G greater than 5.64 velocity and temperature increases without limit as Grundfest (1963) mentioned in his work. Our velocity and temperature profiles agree very well with those included in figure 12 of Bechtel et al. for E=0 (constant viscosity), 3 and 10. However for case E=37 we did not succeed to get a solution. For this case velocity and temperature increases without limit. This means that the flow with Br=2 and E=37, given by Bechtel et al. (2004), does not exist because the characteristics of this flow exceed some criterion. Using a trial and error procedure we found that for Br=2 solutions exist only for E≤23.7. In conclusion, this very interesting work by Bechtel et al. (2004), should be modified with the inclusion of the suitable criteria that are valid for each case treated in their work.